\documentclass[12pt]{iopart}
\usepackage{graphicx}

\begin{document}
\title{Entanglement Entropy for Disjoint Subsystems in XX Spin Chain}

\author{B.-Q. Jin$^\star$ and V. E. Korepin$^\dag$}

\address{$^\star$\ College of Physics and Electronic Information Engineering, Wenzhou University, Wenzhou 325035, P.R. China}
\address{$^\dag$\ C.N.\ Yang Institute for Theoretical Physics, State
University of New York at Stony Brook, Stony Brook, NY 11794-3840,
USA}

\ead{\mailto{jinbq@wzu.edu.cn},\mailto{korepin@insti.physics.sunysb.edu}}

\begin{abstract}
Fisher-Hartwig formula has been successful applied to describe the von Neumann  and R\'enyi entropies of a block of spins in the ground state of XX spin chain. It was based on a determinant representation.
In this paper, we generalize the free fermion method to obtain an exact formulation for the entropy of any finite subsystem in XX spin chain. Based on this, we derive a determinant representation of the entropy of multiple disjoint intervals in the ground state of $XX$ model.
\end{abstract}
\submitto{\JPA}
\maketitle

\section{Introduction}
There is great interest to quantify the entanglement in extended systems in recent years since its ability to detect the scaling behavior in proximity of quantum critical points. Both von Neumann entropy and R\'enyi entropy play a key role in these developments and are so-called entanglement entropies as the measures of entanglement. In the earlier studies of quantum spin chain models, one usually divides the system into two subsystems with each only containing contiguous lattice sites and finds the entanglement entropies of any subsystem when the whole system is in a pure quantum state   \cite{vidal,jin,jin1,kor,calabrese,keating1}.  Recently, there is also growing interest to find the entropy of a non-contiguous subsystem motivated by the search for phase-transition indicators   \cite{pasquier,calabrese1,keating,facchi,wichterich,chen,caraglio} and its relation to conformal field theory   \cite{kor,pasquier,calabrese1}.

The physical system we consider is $XX$ quantum spin chain with lattice index $\{1,2,\cdots \}$. The Hamiltonian for $XX$ quantum spin chain can be written as
\begin{equation}
H=-\sum_{n=1}^{\infty}
\sigma^x_{n}\sigma^x_{n+1}+\sigma^y_{n}\sigma^y_{n+1}
+ h\sigma^z_{n}. \label{xxh}
\end{equation}
All these lattice sites are separated into two sets denoted by ${\cal L}_A$ and ${\cal L}_B$ respectively and we consider
the entropy of subsystem A of spins located on set ${\cal L}_A$ when the spin chain in the ground state.

Following    \cite{Lieb}, we define two sets of Majorana operators
\begin{equation}
c_{2l-1}= \left(\prod_{n<l} \sigma^z_{n}\right) \sigma^x_l~~~\textrm{and}
~~~c_{2l}= \left(\prod_{n<l} \sigma^z_{n}\right) \sigma^y_l\label{mjo1}
\end{equation}
on each site of the spin chain and
\begin{equation}
\tilde{c}_{2l-1}= \left(\prod_{n<l,n\in {\cal L}_A} \sigma^z_{n}\right) \sigma^x_l~~~\textrm{and}
~~~\tilde{c}_{2l}= \left(\prod_{n<l,n\in {\cal L}_A} \sigma^z_{n}\right) \sigma^y_l \label{mjo2}
\end{equation}
on each site of set ${\cal L}_A$.
With the help of the first set of Majorana operators defined by  (\ref{mjo1}), $XY$ quantum spin chain has been exactly solved   \cite{Lieb,mccoy} and following correlations have been obtained
\begin{equation}
\langle GS| c_m|GS\rangle=0,\qquad \langle GS|c_m c_n|GS\rangle =\delta_{mn}+\rmi\,   (\mathbf{B}_L)_{mn}.\label{cccor}
\end{equation}
Here $|GS\rangle$ denotes the ground state of system and  matrix $\mathbf{B}_L$ can be written in a block form as
\begin{equation}
\mathbf{B}_L=\left( \begin{array}{cccc}
\Pi_0 &\Pi_{-1}& \ldots &\Pi_{1-L}\\
\Pi_{1}& \Pi_0&   &   \vdots\\
\vdots &      & \ddots&\vdots\\
\Pi_{L-1}& \ldots& \ldots& \Pi_0
\end{array}     \right) \nonumber
\end{equation}
with
\begin{equation}
\Pi_l =\left( \begin{array}{cc}
               0& g_l\\
               -g_{-l}&0
               \end{array} \right), \quad g_{l}=\frac{1}{2\pi} \int_{0}^{2\pi} \, \rmd \theta\,
e^{-\rmi l \theta} g(\theta).\label{gldef}
               \end{equation}
  For $XX$ model,
  
 \begin{equation} 
 \label{gdef}
 g(\theta)=\cases{
1 & for  $-k_F < \theta <k_F$\\
-1 & for  $k_F < \theta < (2\pi-k_F)$
 \\} \end{equation}
and $k_F=\arccos(|h|/2)$.  Therefore,  we have
\begin{equation}
\langle GS|c_{2m-1} c_{2n}|GS\rangle =-\langle GS|c_{2n} c_{2m-1}|GS\rangle =\rmi\,   g_{m-n}=\rmi\,   g_{n-m}.
\end{equation}

To find the entropy of spins located on set ${\cal L}_A$, the essential step is to find the reduced density matrix for these spins. From   \cite{jin}, one can express
this reduced density matrix with the second set of Majorana operators defined by  (\ref{mjo2}) and their multiplication terms.
Similarly as suggested by    \cite{vidal}, if one can find a set of fermion operators $\{\tilde{b}_i,\tilde{b}^+_i \}$  which is linearly
equivalent to the set of $\{\tilde{c}_j\}$ (i.e. $\tilde{b}_i$ and $\tilde{b}^+_j$ are linear combination of Majorana operators $\{\tilde{c}_j\}$ and transform matrix is of full rank) to satisfy
\begin{equation}
\langle GS| \tilde{b}_i |GS\rangle=\langle GS| \tilde{b}^+_i|GS\rangle=0,\label{defb3}
\end{equation}
\begin{equation}
\langle GS| \tilde{b}_i\, \tilde{b}_j|GS \rangle= \langle GS| \tilde{b}^+_i\, \tilde{b}^+_j|GS \rangle=0\label{defb2}
\end{equation}
and
\begin{equation}
\langle GS| \tilde{b}_i\,  \tilde{b}^+_j|GS \rangle= \langle GS|\tilde{b}_i\, \tilde{b}^+_i|GS\rangle\, \delta_{i,j},\label{defb1}
\end{equation}
then one can find the reduced density matrix
\begin{equation}
\rho_A= \prod_{i\in {\cal L}_A} \Bigl( \langle GS|\tilde{b}^+_i\, \tilde{b}_i|GS
\rangle \tilde{b}^+_i\, \tilde{b}_i+\langle GS|\tilde{b}_i\, \tilde{b}^+_i|GS \rangle \tilde{b}_i\, \tilde{b}^+_i
\Bigr).\label{dmf}
\end{equation}
The reduced density matrix for some specific subsystems of $XY$ model has also been discussed in \cite{peschel1,calabrese1,peschel2}.

In the next section, we will first try to find this set of fermion operators  $\{\tilde{b}_i,\tilde{b}^+_i \}$ satisfying (\ref{defb3}), (\ref{defb2}) and (\ref{defb1}), then obtain the reduced density matrix  and its eigenvalues, finally get the von Neumann entropy and R\'enyi entropy for any finite subsystem.

\section{Exact formulation for entropy of any finite subsystem}

In order to find (or to find whether there exists) a set of fermion operators $\{\tilde{b}_i,\tilde{b}^+_i \}$ satisfying (\ref{defb3}), (\ref{defb2}) and (\ref{defb1}) as done in    \cite{vidal,jin,jin1}, one has to find all correlations of  $\langle GS|\tilde{c}_m\, \tilde{c}_n|GS\rangle$.
In order to find these correlations, we first express Majorana operators $\{\tilde{c}_j\}$ in terms of Majorana operators $\{c_j\}$. From definitions (\ref{mjo1}) and (\ref{mjo2}), we have $\sigma^z_{n}=-\rmi\, c_{2n-1}c_{2n}$ and
\begin{equation}
\tilde{c}_{2l-1}= (\prod_{n<l,n\in {\cal L}_B} \sigma^z_{n}) c_{2l-1}, \qquad\tilde{c}_{2l}= (\prod_{n<l,n\in {\cal L}_B} \sigma^z_{n}) c_{2l}.
\end{equation}
Therefore,
\begin{equation}\fl
\tilde{c}_{2l-1}= \prod_{n<l,n\in {\cal L}_B} (-\rmi\, c_{2n-1}c_{2n} ) c_{2l-1}~~~\textrm{and}
~~~\tilde{c}_{2l}= \prod_{n<l,n\in {\cal L}_B} (-\rmi\, c_{2n-1}c_{2n} ) c_{2l}.
\end{equation}
With the help of Wick Theorem and  (\ref{cccor}), we have
\begin{equation}\fl\quad \qquad
\langle GS|\tilde{c}_m\, \tilde{c}_n|GS\rangle= \langle GS|c_m\, \prod_{k\in {\cal L}_B \cap {\cal L}_{\left[(m+1)/2\right]\left[(n+1)/2\right]} }\left[ -\rmi\, c_{2k-1}c_{2k}\right]\, c_n|GS\rangle,
\end{equation}
where ${\cal L}_{mn}$ denotes the set of lattice sites between (but not including) $m$-th and $n$-th sites, and $[x]$ means the integer part of $x$. Therefore, set ${\cal L}_{\left[(m+1)/2\right]\left[(n+1)/2\right]}$ includes all lattice sites  between (but not including)
 lattice sites on which Majorana operators $\tilde{c}_m$ and $\tilde{c}_n$ are defined.
The right-hand side of the above identity for correlations  $\langle GS|\tilde{c}_m\, \tilde{c}_n|GS\rangle$ can be further expressed by correlations  $\langle GS|c_p c_q|GS\rangle$ according to Wick Theorem. Let's assume there is $K$ lattice sites inside of the set ${\cal L}_B \cap {\cal L}_{\left[(m+1)/2\right]\left[(n+1)/2\right]}$  with lattice index $\{D_1,D_2,\cdots,D_K\}$, then by brute force one can find
\begin{equation}
 \langle GS|\tilde{c}_{2p-1}\, \tilde{c}_{2q}|GS\rangle
=\rmi\, \det(T_{pq})\label{me2}
\end{equation}
and \begin{equation}
 \langle GS|\tilde{c}_{2p}\, \tilde{c}_{2q}|GS\rangle=\langle GS|\tilde{c}_{2p-1}\, \tilde{c}_{2q-1}|GS\rangle=\delta_{pq}.\label{twopoints2}
\end{equation}
Here $(K+1)\times (K+1)$ matrix $T_{pq}$ is Toeplitz-like defined with row index  $i\in \{p, D_1,D_2,\cdots,D_K\}$, column index  $j\in \{q,D_1,D_2,\cdots,D_K\}$,
and matrix element in $i$-th row and $j$-th column by
 \begin{equation}\left(T_{pq}\right)_{ij}=g_{i-j}= \frac{1}{2\pi} \int_{0}^{2\pi} \, \rmd \theta\,
e^{-\rmi (i-j) \theta} g(\theta) \label{toeplitzlike} \end{equation}
and $g(\theta)$  in  (\ref{gdef}). Interchanging row and column of $T_{pq}$ and noticing $g_l=g_{-l}$ for $XX$ spin chain, one can find that \begin{equation}\det(T_{pq})=\det(T_{qp}).\label{tsym}\end{equation}
Now let us define
\begin{equation}\tilde{a}_{l}=\frac{1}{2}\left( \tilde{c}_{2l-1}-\rmi\, \tilde{c}_{2l}\right)\quad \textrm{then} \quad \tilde{a}^+_{l}= \frac{1}{2}\left(\tilde{c}_{2l-1}+\rmi\, \tilde{c}_{2l}\right) \quad \textrm{if $l\in {\cal L}_A$} \label{def-al}.\end{equation}
From  (\ref{onepoint0}), we have
\begin{equation}\langle GS|\tilde{a}_{l}|GS\rangle=\langle GS|\tilde{a}^{+}_{l}|GS\rangle=0.\label{onepoint1}\end{equation}
Let us define matrix $\mathbf{{A}}$   \cite{Eisler} through
\begin{equation}\langle GS| \tilde{a}_{m}\, \tilde{a}^+_{n}|GS\rangle\equiv \frac{\delta_{mn}}{2} + \frac{\left(\mathbf{{A}}\right)_{mn}}{2}. \end{equation}
Then from  (\ref{twopoints2}) we can find that
\begin{equation}
\left(\mathbf{{A}}\right)_{mn}=\frac{\rmi}{2}\left( \langle GS|\tilde{c}_{2m-1}\, \tilde{c}_{2n}|GS\rangle+ \langle GS|\tilde{c}_{2n-1}\, \tilde{c}_{2m}|GS\rangle \right).\label{me1}
\end{equation}
From  (\ref{me2}) and (\ref{tsym}), we obtain that
\begin{equation}\left(\mathbf{{A}}\right)_{mn} =-\frac{1}{2}\left[\det(T_{mn})+\det(T_{nm})\right]=-\det(T_{mn}). \end{equation}
  Hence, in $XX$
quantum spin chain, $\mathbf{{A}}$ is a real symmetric matrix with \begin{equation}\left(\mathbf{{A}}\right)_{mn} =-\det(T_{mn}).\label{adiagonal} \end{equation}
Here $T_{mn}$ is Toeplitz-like matrix defined in  (\ref{toeplitzlike}).
Similarly, we have\begin{equation}
 \langle GS|\tilde{a}_{l}\, \tilde{a}_{m}|GS\rangle=\langle GS\left|\frac{\tilde{c}_{2l-1}-\rmi\, \tilde{c}_{2l}}{2}\, \frac{\tilde{c}_{2m-1}-\rmi\, \tilde{c}_{2m}}{2}\right|GS\rangle=0.\label{twopoints}
\end{equation}
\begin{equation}
 \langle GS|\tilde{a}^+_{l}\, \tilde{a}^+_{m}|GS\rangle=\langle GS\left|\frac{\tilde{c}_{2l-1}+\rmi\, \tilde{c}_{2l}}{2}\, \frac{\tilde{c}_{2m-1}+\rmi\, \tilde{c}_{2m}}{2}\right|GS\rangle=0.\label{twopoints0}
\end{equation}
Now let us take an example, in which set ${\cal L}_A$ contains the lattice site $1$ and lattice site $3$, and work out its matrix $\mathbf{A}$ explicitly. For this case,  $\mathbf{A}$ is a two by two matrix with row (and column) index taking $\{1,3\}$, i.e.
\begin{equation}
\mathbf{A}=\left( \begin{array}{cc}
{A}_{11} &{A}_{13} \\
{A}_{31} &{A}_{33}
\end{array}     \right).
\end{equation}
All matrix elements of above matrix can be obtained by  (\ref{adiagonal}). We have
\begin{equation}{A}_{11}={A}_{33}=-g_{0}. \end{equation}
We also know that matrix element ${A}_{13}$ is related to the matrix $T_{13}$, which is defined in  (\ref{toeplitzlike}) with row index taking $\{1, 2\}$ and column index taking $\{3,2\}$.  Hence, we have
\begin{equation} T_{13}=\left( \begin{array}{cccc}
g_{-2} &g_{-1} \\
g_{-1} &g_{0}
\end{array}     \right)=\left( \begin{array}{cccc}
g_{2} &g_{1} \\
g_{1} &g_{0}
\end{array}     \right).  \end{equation}
Therefore, we obtain
\begin{equation}{A}_{13}={A}_{31}=g_1^2-g_2\, g_0. \end{equation}

Since matrix $\mathbf{{A}}$  is a real symmetric matrix, it's diagonalizable with an orthogonal transformation, i.e.  there exists an orthognal matrix $V$ satisfying
\begin{equation}V \mathbf{A} V^{T}=\mathbf{A}_{d}, \end{equation}
where $\mathbf{A}_{d}$ is diagonal matrix.  Let's denote
$\left(V\right)_{mn}=v_{mn}$, $\left(\mathbf{A}_{d}\right)_{mn}=\nu_m\delta_{mn}$ and define $\tilde{b}_i=\sum_m v_{im} \tilde{a}_{m}$, then we have  $\tilde{b}_i^+=\sum_m v_{im} \tilde{a}_{m}^+$. Therefore,
\begin{equation}\langle GS|\tilde{b}_i\,  \tilde{b}^+_j|GS \rangle=\sum_{m,n} v_{im}\langle GS| \tilde{a}_{m} \tilde{a}^{+}_{n}|GS\rangle v_{jn}= \frac{1+\nu_i}{2}\, \delta_{ij}.\end{equation}
Similarly, from  (\ref{twopoints}) and (\ref{twopoints0}), we have \begin{equation}\langle GS|\tilde{b}_i\,  \tilde{b}_j|GS \rangle=\langle GS|\tilde{b}^+_i\,  \tilde{b}^+_j|GS \rangle=0.\end{equation}
With the help of Wick Theorem and  (\ref{cccor}), we also have
\begin{equation}\langle GS|\tilde{c}_{m}|GS\rangle=0.\label{onepoint0}\end{equation}
Therefore,
\begin{equation}
\langle GS| \tilde{b}_i |GS\rangle=\langle GS| \tilde{b}^+_i|GS\rangle=0.\end{equation}
Thus we find a set of fermion operators  $\tilde{b}_i$ and $\tilde{b}^+_j$ satisfying   (\ref{defb3}), (\ref{defb2}) and (\ref{defb1}) through the diagonalization of matrix $\mathbf{A}$ and obtain
\begin{equation}\langle GS|\tilde{b}_i\,  \tilde{b}^+_i|GS \rangle=\frac{1+\nu_i}{2},  \quad  \langle GS|\tilde{b}^+_i\, \tilde{b}_i|GS \rangle=\frac{1-\nu_i}{2},\end{equation}
where $\{\nu_i\}$ are the eigenvalues of matrix $\mathbf{{A}}$.
Once we get this set of $\{\nu_i\}$, the von Neumann entropy of subsystem $A$ can be expressed as
\begin{equation}S(\rho_A)=\sum_i\left[-\frac{1+\nu_i}{2} \ln \frac{1+\nu_i}{2}-\frac{1-\nu_i}{2} \ln
\frac{1-\nu_i}{2}\right],\label{vons}\end{equation}
and R\'enyi entropy
\begin{equation}S_{\alpha}(\rho_A)=\sum_i\frac{1}{1-\alpha}\ln \left[\left(\frac{1+\nu_i}{2}\right)^{\alpha}+
\left(\frac{1-\nu_i}{2}\right)^{\alpha}\right],\quad \alpha\neq 1\textrm{~and~}\alpha>0. \label{renyis}\end{equation}

The way presented above can be used to exactly albeit numerically determine the entropy of system A containing several disjoint intervals.

\section{Determinant representation for entropy of any subsystem}
\begin{figure}[htb]
\begin{center}
\includegraphics[width=2.5in,clip]{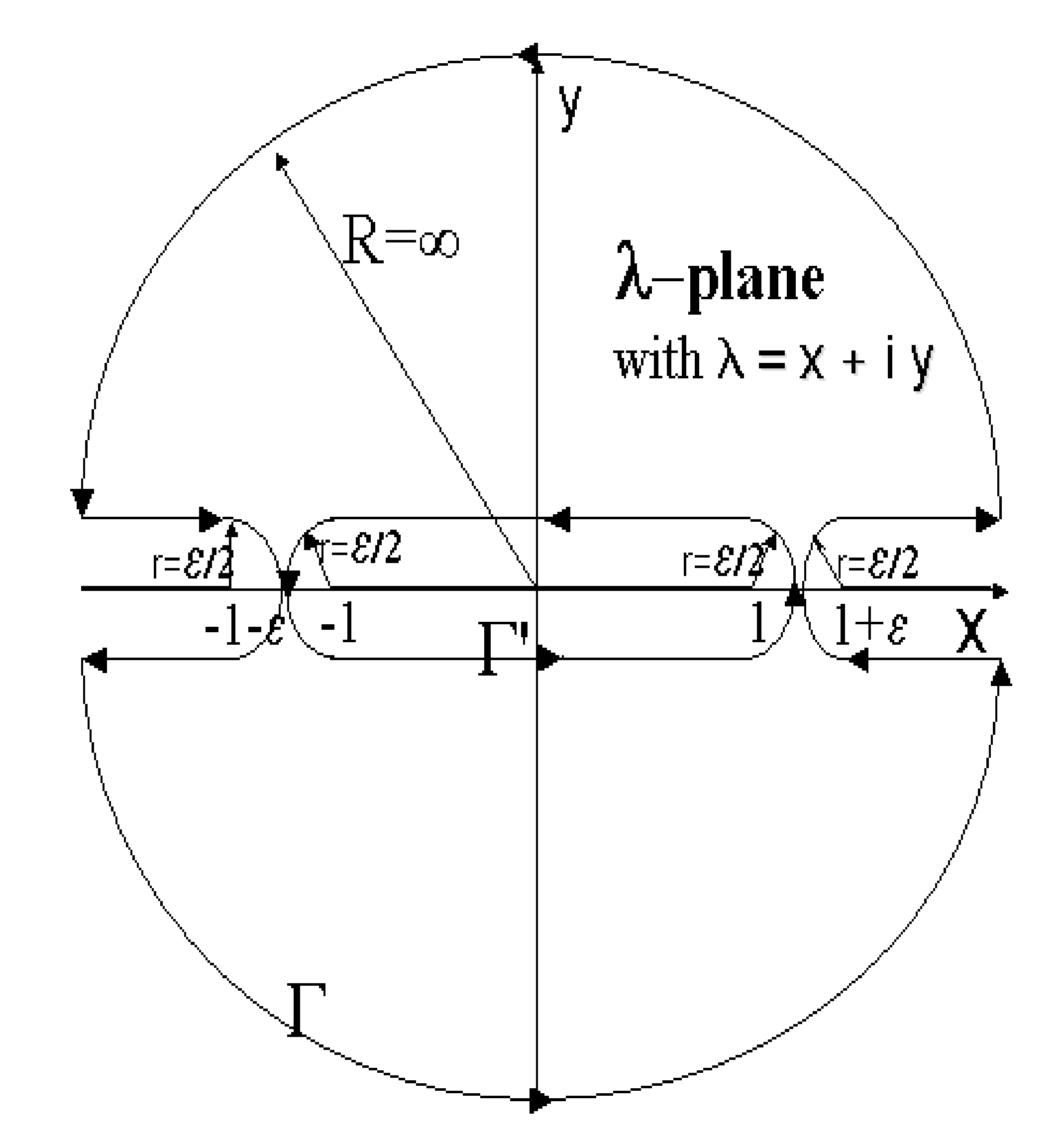}
\end{center}
\caption{\it Contours \mbox{$\Gamma'$} (smaller one) and
\mbox{$\Gamma $} (larger one). Bold lines $(-\infty, -1-\epsilon)$
and $(1+\epsilon,\infty)$ are the cuts of the integrand
$e(1+\epsilon,\lambda)$. Zeros of $D_{L}(\lambda)$
 are located on the bold line $(-1, 1)$. The
arrows indicate the directions of integrations, and $\mathrm{r}$
and $\mathrm{R}$ are the radius of the circles. $\P $  }
\label{fig1}
\end{figure}

In order to develop a way of  analytical treatment as in  \cite{jin}, let us define
\begin{equation} \mathbf{\tilde{A}}\equiv  \lambda\, I- \mathbf{A},\quad D(\lambda)\equiv\det{\mathbf{\tilde{A}}}, \label{def-atilde} \end{equation} where $I$ is the identity matrix.
Then we will have the determinant representation for the von Neumann entropy of subsystem $A$
 \begin{equation}
S(\rho_A)=\lim_{\epsilon \to 0^+} \frac{1}{2\pi \rmi}
\oint_{\Gamma'} \rmd \lambda\,  e(1+\epsilon, \lambda)
\frac{\rmd}{\rmd \lambda} \ln
D(\lambda)\label{eaa}
\end{equation}
with \begin{equation}
 e(x, \nu)= -\frac{x+\nu}{2} \ln \frac{x+\nu}{2}-\frac{x-\nu}{2} \ln
\frac{x-\nu}{2}
\end{equation}
and R\'enyi entropy
 \begin{equation}
S_{\alpha}(\rho_A)=\lim_{\epsilon \to 0^+} \frac{1}{2\pi \rmi}
\oint_{\Gamma'} \rmd \lambda\,  e_{\alpha}(1+\epsilon, \lambda)
\frac{\rmd}{\rmd \lambda} \ln
D(\lambda)\label{eab}
\end{equation}
with \begin{equation}
 e_{\alpha}(x, \nu)= \frac{1}{1-\alpha}\ln\left[\left( \frac{x+\nu}{2}\right)^{\alpha}+ \left(
\frac{x-\nu}{2}\right)^{\alpha}\right].
\end{equation}
Here the contour \mbox{$\Gamma'$} is depicted in \Fref{fig1},
which encircles all zeros of \mbox{$D(\lambda)$}.

For the case of set ${\cal L}_A$ only containing contiguous sites  \cite{vidal}, $\mathbf{\tilde{A}}$  can be expressed as Toeplitz matrix with Toeplitz operators
\begin{equation}{\cal
G}_A(\theta)=\lambda+g(\theta) \label{toeplitzgen}\end{equation}
 where $g(\theta)$ is defined in  (\ref{gdef}). With the help of Weiner-Hopf factorization and Fisher-Hartwig conjecture, integeration of  (\ref{eaa}) and (\ref{eab}) has been explicitly worked out for this case \cite{jin}.

Although it's harder to give an explicit expression for $\mathbf{\tilde{A}}$ in the case of set ${\cal L}_A$ containing non-contiguous sites,  $\mathbf{\tilde{A}}$ is of the form
\begin{equation}
\mathbf{\tilde{A}}=\left( \begin{array}{cccc}
\mathbf{\tilde{A}}_{n_1} &\ldots & \ldots &\ldots\\
\ldots & \mathbf{\tilde{A}}_{n_2}& \ddots  &   \vdots\\
\vdots &  \ddots    & \ddots&\vdots\\
\ldots & \ldots& \ldots& \mathbf{\tilde{A}}_{n_m}
\end{array}     \right),\label{defga}
\end{equation}
with row (and column) index taking $1,2,\cdots, n_1-1,n_1, n_1+k_1+1,n_1+k_1+2,\cdots, n_1+k_1+n_2, n_1+k_1+n_2+k_2+1, \cdots$, if the first $n_1$ lattice sites belong to ${\cal L}_A$, then the next $k_1$ lattice sites belong to ${\cal L}_B$, then the next $n_2$ lattice sites belong to ${\cal L}_A$,  and so on. These $n_{i}\times n_{i}$ sub-matrix  $\mathbf{\tilde{A}}_{n_{i}}$ are Toeplitz matrix with generator ${\cal
G}_A(\theta)$ defined in  (\ref{toeplitzgen}), and the elements of off-diagonal block $\left(\mathbf{\tilde{A}}\right)_{ij}=-\left(\mathbf{{A}}\right)_{ij}$ related to the determinant of Toeplitz-like matrix as seen in  (\ref{adiagonal}). So far, we got the determinant representation for the subsystem $A$ containing several disjoint intervals. To have analytical treatment further, it will be necessary to tackle with the determinant of Toeplitz-like matrix  $T_{mn}$  defined in  (\ref{toeplitzlike}).

\section{An application to two disjoint interval subsystem}
Now let us come to a simple but non-trivial case  \cite{pasquier}, in which subsystem $A$ contains two disjoint  intervals with  with  equal length $m$ and distance $m$, i.e. lattice sites with index $\{1,2,\cdots, m\}$ (denoted as  $A_{1}$) and $\{2m+1,2m+2,\cdots,3m\}$ (denoted as $A_{2}$) belong to subsystem $A$, but lattice sites with index $\{m+1,m+2,\cdots,2m\}$ do not belong to subsystem $A$. We also take the external field $h=0$ in the model Hamiltonian  (\ref{xxh}). Therefore, the ground state of this system is in critical phase.  We want to find the mutual information   \cite{pasquier,calabrese1} between these two intervals, which is defined as
\begin{equation}I_{A_{1}:A_{2}}=S_{A_{1}}+S_{A_{2}}-S_{A}, \end{equation}
when the system is in the ground state.

From (\ref{adiagonal}),
we have the matrix $\mathbf{A}$ for subsystem $A$ expressed as a block matrix
\begin{equation}
\mathbf{A}=\left( \begin{array}{cccc}
\mathbf{A}_{1,1} &\mathbf{A}_{1,2} \\
\mathbf{A}_{2,1} &\mathbf{A}_{2,2}
\end{array}     \right).\label{samatrix} \end{equation}
Here related blocks can be expressed as
\begin{equation}
\mathbf{A}_{1,1}=\mathbf{A}_{2,2}=\left( \begin{array}{cccc}
-g_0 &-g_{-1}& \ldots &-g_{1-m}\\
-g_{1}& -g_0&   &   \vdots\\
\vdots &      & \ddots&\vdots\\
-g_{m-1}& \ldots& \ldots& -g_0
\end{array}     \right),\label{sa1matrix} \end{equation}
\begin{equation}
\mathbf{A}_{1,2}=\mathbf{A}_{2,1}^{T}=\left( \begin{array}{cccc}
{\cal A}_{11} &{\cal A}_{12}& \ldots &{\cal A}_{1m}\\
{\cal A}_{21}& {\cal A}_{22}&   &   \vdots\\
\vdots &      & \ddots&\vdots\\
{\cal A}_{m1}& \ldots& \ldots& {\cal A}_{mm}
\end{array}     \right)
\end{equation}
with
\begin{equation}
{\cal A}_{ij}=-\left| \begin{array}{ccccc}
g_{i-j-2m} &g_{i-m-1} &g_{i-m-2}& \ldots &g_{i-2m}\\
g_{1-j-m}& g_{0}&g_{-1}& \ldots  &   g_{1-m}\\
g_{2-j-m}& g_{1}&g_{0}&   &   \vdots\\
\vdots & \vdots  &   & \ddots&\vdots\\
g_{-j}&g_{m-1}& \ldots& \ldots& g_{0}
\end{array}     \right|\end{equation}
and from  (\ref{gldef})
\begin{equation}g_{j}=\frac{1-(-1)^{j}}{j\pi}\, \sin\left(\frac{j\pi}{2}\right). \end{equation}

\begin{figure}[htb]
\begin{center}
\includegraphics[width=2.5in,clip]{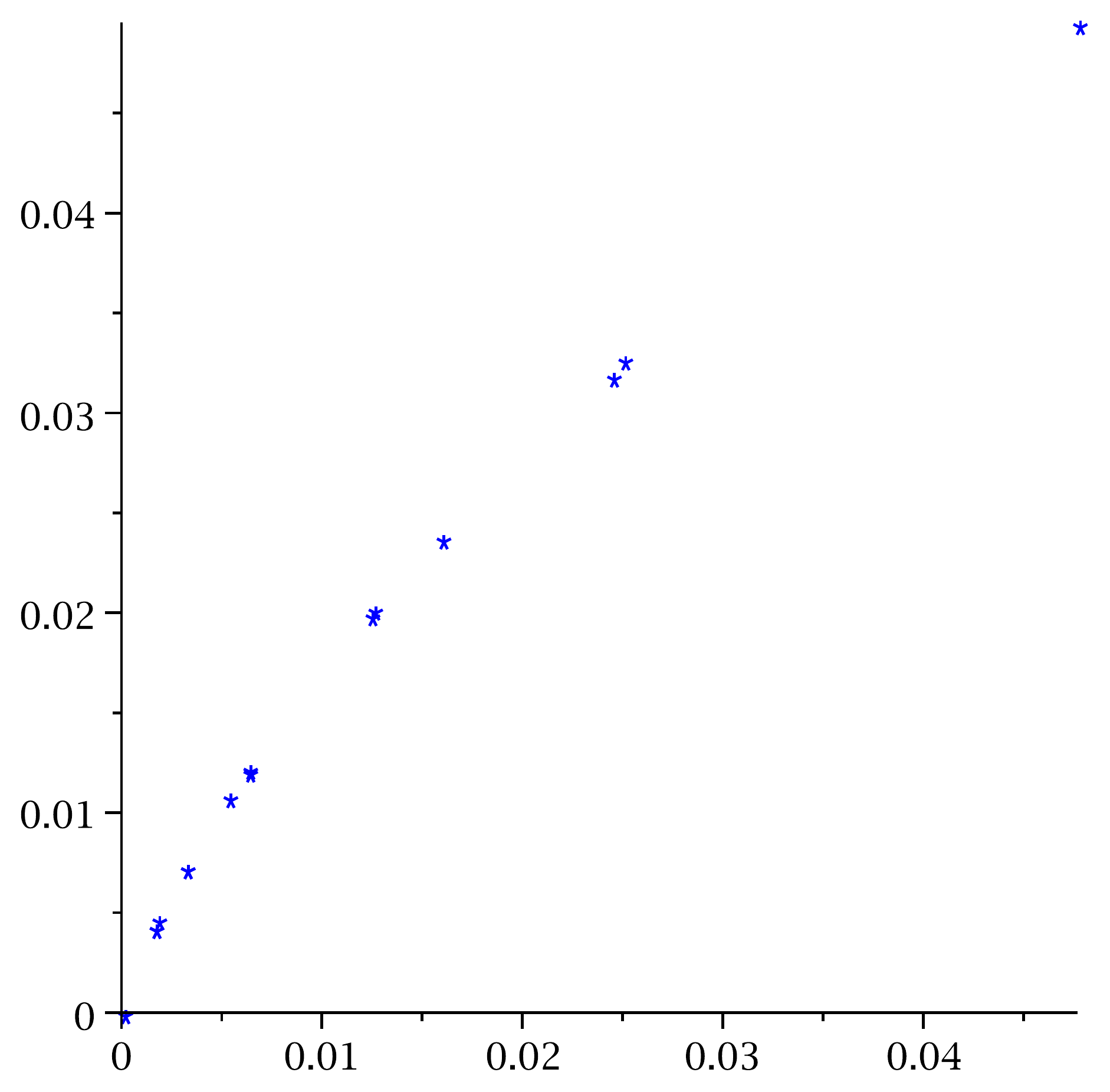}
\end{center}
\caption{\it  $I_{A_{1}:A_{2}}$($y$-axis) as function of $1/m$ ($x$-axis). All points except $(0,0)$ are calculated by taking $m=21,40,41,63,80,81,160,161,189,320,321,567,640,641$ respectively. Here,$I_{A_{1}:A_{2}}$ is the mutual information between two intervals, $m$ is the length of intervals and also the distance between two intervals.  $\P $  }
\label{fig2}
\end{figure}

Finding all eigenvalues of matrix defined in  (\ref{samatrix}) and substituting them into  (\ref{vons}), one can find $S_{A}$. Similarly, one can obtain $S_{A_{1}}=S_{A_{2}}$ by finding all eigenvalues of matrix defined in   (\ref{sa1matrix}) and substituting them into  (\ref{vons}).
The mutual information between these two intervals is presented in \Fref{fig2}.
From \Fref{fig2}, we find that the mutual information between two intervals will vanishes possibly when both the length of intervals and the distance between two intervals goes into infinite in the same scale even the system is in critical phase. It coincides the one obtained in \cite{pasquier} in the case that the ratio of interval length to chain length goes to zero.

\section{Summary}
We have generalized the numerical calculation method for the entropy of one interval, which first appeared in  \cite{vidal},  to the case of multiple disjoint intervals in the ground state of $XX$ model. We use this formulation to find the mutual information between two intervals with the separation and the length of interval same. We confirm that the mutual information will vanishes possibly when the length scale goes into infinite even the system is in critical phase. We also derive the determinant representation of the entropy of multiple disjoint intervals for further analytical treatment.

\ack  We would like to thank Prof. J.\,H.\,H. Perk, Prof. A.\,R. Its and Prof. B.\,M. McCoy for useful discussions. This work was partially supported by the National Science Foundation (USA) under Grant DMS-0905744.

\end{document}